\documentclass[12pt]{article}
\usepackage{geometry}                
\usepackage{graphicx}
\usepackage{amssymb}
\usepackage{amsmath}
\usepackage{amsfonts}
\usepackage{graphicx}
\usepackage{epstopdf}
\usepackage{color}
\usepackage{cite}

\newcommand{\be}{\begin{equation}}
\newcommand{\ee}{\end{equation}}
\newcommand{\bea}{\begin{eqnarray}}
\newcommand{\eea}{\end{eqnarray}}
\newcommand{\beas}{\begin{eqnarray*}}
\newcommand{\eeas}{\end{eqnarray*}}

\newcommand{\w}{\vec{w}}

\newcommand{\xk} {{\vec x}_{\tau}}
\newcommand{\wxk} {{\vec w} \cdot \xk }

\title{Optimal Liquidation Strategies Regularize Portfolio Selection}
\author{Fabio Caccioli,\\ 
{\em Santa Fe Institute, 1399 Hyde Park road, 87501 NM, USA} \and Susanne Still,\\ {\em Information and Computer Sciences, University of Hawai`i at}\\ {\em M\={a}noa, 1680 East-West Road, Honolulu 96822, Hawai`i, USA} \and Matteo Marsili,\\ {\em Abdus Salam International Centre for Theoretical Physics,}\\ {\em Strada Costiera 11, 34151 Trieste, Italy} \and Imre Kondor,\\{\em Collegium Budapest - Institute for Advanced Study and Department of} \\{\em Physics of
Complex Systems, E\"{o}tv\"{o}s University, Budapest, Hungary}}

\begin{document}
\maketitle
\begin{abstract}

We consider the problem of portfolio optimization in the presence of market impact, and derive optimal liquidation strategies. We discuss  in detail the problem of finding the optimal portfolio under Expected Shortfall (ES) in the case of linear market impact. We show that, once market impact is taken into account, a regularized version of the usual optimization problem naturally emerges. We characterize the typical behavior of the optimal liquidation strategies, in the limit of large portfolio sizes, and show how the market impact removes the instability of ES in this context. 

\end{abstract}

\section{Introduction}

The optimization of large portfolios is known to be highly unstable, due to large sample to sample fluctuations. This instability arises from the fact that risk measures need to be estimated from empirical observations in the market and for large portfolios we can never have enough data: by the very nature of portfolio selection, the sampling frequency cannot be high, and the look-back period cannot be too long. Therefore, the length $T$ of the available time series cannot be sufficiently large compared to the number $N$ of the different items in the portfolio: for large portfolios $T$ may be, at best, of the same order of magnitude as $N$.  Clearly, classical statistical methods cannot be expected to work in this regime. This difficulty has been well known in portfolio theory, and a host of methods have been put forward to handle it \cite{eltongruber,jobson1979,jorion1986,frost_savarino,safePO,Jagannathan2003,LedoitWolf2003, LedoitWolf2004,LedoitWolfHoney,DeMiguel2007,Garlappi2007,Golosnoy2007,Kan2007,frahm_memmel,DeMiguel2009,Laloux1999,Plerou1999,Laloux2000,Plerou2002,burda,Potters2005}, but the problem was put in a particularly sharp light when it was recognized that for a critical value of the ratio $N/T$ the estimation error actually diverges \cite{pafka2003,kondor2007}. 

A similar statistical problem arises in many other areas, and methods known as regularization can be used as a remedy, see \cite{SK} and references therein. In the context of portfolio selection, regularization penalizes large excursions of the portfolio weight vector by imposing a constraint on the length, as measured in terms of a suitably chosen norm \cite{SK}. Regularized portfolio optimization, in general, then has two choices to make \cite{SK}: (i) which risk function to use, and (ii) which Lp-norm to use as a regularizer. 

Regularized portfolio optimization was studied in \cite{SK} with the increasingly popular risk measure Expected Shortfall (ES), and the L2-norm. This scheme was shown in \cite{SK} to be related to support vector regression \cite{CortesVapnik95, Nu-SVM}, and \cite{SK} introduced the necessary modifications to the support vector machine algorithm which are required by the asymmetry of the ES risk measure and by the extra constraint imposed by the fixed budget. A number of other filtering techniques that use the variance as a risk measure also implement regularization \cite{safePO, Brodie09}, even when the link to statistical learning theory is not made 
explicit.\footnote{This includes works related to covariance shrinkage 
\cite{Jagannathan2003,LedoitWolf2003,LedoitWolf2004,LedoitWolfHoney,DeMiguel2009}. For a discussion see also \cite{SK} and references therein.}

The choice of the risk measure is a subject of ongoing debate \cite{Szego}. In this paper we focus on the optimization of portfolios under the Expected Shortfall. ES is the mean loss above a high quantile.\footnote{Note the sign convention: in the context of risk measures losses are counted positive.} The reason for our choice is that ES provides a more accurate estimate of risk when returns have fat tailed distributions \cite{mandelbrot1963,fama1965,muller1990,hols1991,mantegna1995} and it gives a more faithful representation of large losses than Value at Risk (VaR)  \cite{jorion,riskmetrics} that can be identified by the quantile itself. In addition, ES can be computed by fast linear programming algorithms \cite{Rockafellar} and, most significantly, it was shown \cite{acerbi2,acerbi3,acerbi4} to belong to the set of coherent risk measures \cite{Artzner99}. These properties make it an attractive alternative to VaR which has been criticized  for its lack of convexity \cite{Artzner99, embrechts,acerbi1}. 

As for the choice of the regularizer, one has to realize that various regularizers act differently. The L2 norm has a tendency to favor equal weights, and thus acts as a diversification pressure \cite{SK, Bouchaudpotters}. The L1 norm can lead to more sparse solutions, see e.g. \cite{Tibshirani96, Chen}. In portfolio optimization it corresponds to an exclusion of short positions \cite{DeMiguel2009,Brodie09}. We refer the interested reader to \cite{Brodie09}, where its use with the variance as a risk measure is explored.\footnote{L1 can produce a stable solution only if there is real structure in the data. If there is no structure (say all the items are more or less equivalent), then the solution with the L1 norm will still be sparse, but unstable, as the set of zero weights will vary from sample to sample \cite{Kondor4}.} 
It is evident that the proper choice of the regularizer is not merely a matter of mathematical convenience: one has to motivate it on the basis of the nature of the problem at hand. In a Bayesian context, the choice of the regulariser expresses prior information, as it reflects what we know, or believe to know, about the possible structure in the data. 
 
In this paper we suggest that regularization inherently arises from considering the market impact of portfolio liquidations strategies, and we show how a linear assumption for the market impact leads to the L2 norm as a regularizer. 

The fact that liquidity considerations effectively regularize portfolio selection is quite natural, in view of the origin of the instability and of the no-arbitrage hypothesis. In brief, it was shown \cite{hasszan2008,kondor2008} that the instability of risk measures arises because, when portfolio returns are estimated on a data set of finite length, then an {\em accidental arbitrage} may appear, corresponding to a zero cost portfolio which happens to have a positive return on that particular data set. When this occurs, the minimization of risk measures that have no lower bound (like ES, for example) dictates to take an infinitely long position on these portfolios. However, in real markets, we expect that realized prices will react to the liquidation of a portfolio by adjusting prices in such a way as to eliminate accidental arbitrages.  Therefore, taking the market impact of the liquidation into account as part of the portfolio optimization problem should intuitively regularize  portfolio optimization.

The paper is organised as follows. In Section \ref{ESsect} we define the Expected Shortfall risk measure, set up the problem of its optimization, discuss its instability, and recall how to overcome this difficulty by regularization, as suggested in \cite{SK}. 

We show in section \ref{MarketImpact} that regularization can be derived from estimating the risk of portfolio liquidation, within a linear assumption of market impact. This provides a clear financial motivation for choosing the L2-norm as a regularizer in the given problem. Note that this derivation does not depend on the choice of the risk function and that it therefore justifies and motivates the use of an L2-regularizer also in methods that use the variance as a risk measure. This market impact consideration is related to a recent proposal suggesting that portfolio optimization should be defined in terms of liquidation strategies \cite{Acerbi2009}. 

Section \ref{regularize} provides a concrete illustration of how the regularizer works in typical cases. Following \cite{CKM07}, we derive analytic results for typical instances of large portfolios, generated from a simple distribution. This shows that regularization, as suggested in \cite{SK}, indeed removes the instability observed in \cite{CKM07}, but it also shows the non-trivial role played by the confidence level: the more the risk measure is concentrated in the tail of the distribution, the weaker is the regularization induced by liquidity considerations. 

The final section concludes with a short summary. Details of the derivation of the results presented in section \ref{regularize} are relegated to an Appendix.

\section{Expected Shortfall, instability and regularization}
\label{ESsect} 

We consider a portfolio $\{w_1\ldots w_N\}$ of $N$ assets with returns $\{x_i\}$, with $w_i$ the position of asset $i$. We impose a global budget constraint $\sum_{i=1}^N w_i=wN$ and we allow the weights $\{w_i\}$ to take any real value.
The problem we address is that of finding the optimal weights $\{w_i\}$ that minimize the expected shortfall, defined as follows.
Given the loss $l(\{w_i\}|\{x_i\})=-\sum_i w_i x_i$, the probability for such loss to be smaller than a threshold $\alpha$ is
\begin{equation}
P_<(\{w_i\},\alpha)=\int \prod_i dx_i p(\{x_i\}) \theta(\alpha-l(\{w_i\}|\{x_i\})),
\end{equation}
with $\theta(x)=1$ if $x>0$ and $\theta(x)=0$ otherwise.
The associated $\beta{\rm VaR}$ is defined as
\be
\beta{\rm VaR}(\{w_i\})={\rm min}\{\alpha: P_<(\{w_i\},\alpha)\ge \beta\},
\ee 
while the Expected Shortfall ${\rm ES}(\{w_i\})$ is given by
\be
{\rm ES}(\{w_i\})=\frac{1}{1-\beta}\int\prod_i dx_i p(\{x_i\}) l(\{w_i\}|\{x_i\})\theta(l(\{w_i\}|\{x_i\})-\beta{\rm VaR}(\{w_i\})).
\ee
The calculation of the ES \cite{Rockafellar} can be obtained through the minimization of the function 
\be\label{F}
F_{\beta}(\{w_i\},v)=v+\frac{1}{1-\beta}\int \prod_i dx_i p(\{x_i\})[l(\{w_i\}|\{x_i\})-v]^+
\ee
with respect to the auxiliary parameter $v$, where $[x]^+=(x+|x|)/2$, i.e. 
\be
{\rm ES}(\{w_i\})={\rm min}_vF_{\beta} (\{w_i\},v),
\ee
Approximating the integral in (\ref{F}) by sampling the probability distributions of returns, 
the problem can be reduced to the calculation of the minimum of the cost function
$$E[v,\{u_{\tau}\}]=(1-\beta) Tv+\sum_{\tau=1}^T u_{\tau}$$ under the constraints
$$u_{\tau}\ge
0~~\forall\tau,$$ $$u_{\tau}+v+\sum_{i=1}^N x_{i,\tau} w_i\ge 0~~\forall\tau$$ and $$\sum_i w_i=wN,$$ 
where we have introduced a new set of variables defined as 
$$u_{\tau}=\frac{\left(-v-\sum_{i=1}^N w_i x_{i,\tau}\right)+\left|(-v-\sum_{i=1}^N) w_i x_{i,\tau})\right|}{2}.$$

Portfolio optimization often assumes that weights should satisfy a further constraint that fixes the expected return to some target value. However, it has been remarked \cite{Jagannathan2003,merton1980} that estimation error in sample expected returns is so large that nothing much is lost in ignoring this constraint altogether, with no appreciable effect on out-of-sample performance \cite{jorion1986}. Also, in applications such as index tracking, where the objective is to mimic a benchmark portfolio as closely as possible, this constraint is not needed. Finally, the issues we address below \cite{hasszan2008}, the main argument and the results are not affected by the presence of a constraint on expected returns. So, for the sake simplicity, we omit this constraint in what follows.

\subsection{Instability of Expected Shortfall}

Being a conditional average, ES is not bounded from below: if a portfolio produces a large gain, rather than a loss, then ES takes a large negative value. Now, on a finite sample it may happen that one of the items, or a combination of items, {\em dominates} the others, i.e. produces a larger return at each time point than the rest. When such an apparent arbitrage occurs, the optimization of ES suggests to go as long as possible in the dominating asset and correspondingly short in the dominated ones. In particular, if there are no other constraints except for the fixed budget, then this leads to a runaway solution and to a seemingly infinite return. Therefore, for finite $N$ and $T$ the optimization of ES will have a finite solution with a probability always less than one. This probability quickly approaches one as $N/T$ goes to zero, and quickly approaches zero as $N/T$ goes to infinity. The transition between the two limits becomes sharper and sharper as $N$ and $T$ go to infinity such that their ratio is fixed, which is the realistic limit to consider for large institutional portfolios. In this limit there will be a critical value of the ratio $N/T$ where a sharp transition occurs between the region where the optimization of ES leads to a finite solution and the one where it does not. This instability of ES was pointed out in \cite{kondor2007}, where the critical $N/T$ as function of the cutoff beyond which the conditional average is calculated (i.e. the phase diagram) was determined numerically, while \cite{CKM07}  derived an analytic expression for it by the help of tools borrowed from the statistical physics of random systems.  


Although the ES is a specific case of the several possible risk measures, it was explicitly demonstrated \cite{kondor2008} for all the coherent risk measures  \cite{Artzner99} that  the presence of apparent arbitrages constitutes a sufficient condition for the portfolio optimization problem to be unfeasible. 
Moreover, the same instability was shown to characterize also  portfolio optimization problems under downside risk measures \cite{hasszan2008}, including parametric VaR, one of the standard tools in banking.
 In the following, we will then focus on the specific case of ES with the idea that the results derived can be generalized to such wider classes of risk measures.

\subsection{Regularized Portfolio Optimization}

The essence of the problem discussed above, namely that the dimension (here $N$) is too high compared with the size of the statistical sample ($T$), is common to all fields where complex modeling or optimization problems arise. The field of statistical learning theory/machine learning (see e.g. \cite{VapnikCh71, Vapnik95, Vapnik98}) has developed powerful and systematic methods to deal with the difficulty of insufficient data. The insights gained in that area can be directly applied to portfolio optimization, as pointed out in \cite{SK}. 

The main observation is that the instability is caused by over-fitting, which in turn is caused by the fact that the empirical risk is minimized in a regime in which there is not enough data to guarantee small actual risk. The weights $\{w_i\}$ constitute a linear model of the observed returns, and the model capacity has to be constrained in order to avoid over-fitting. The capacity of the linear model is monotonic in the length of the weight vector, so that minimizing that length results in better generalization performance. 

In the context of portfolio selection this means that the resulting portfolio better reflects the actual variations in the data by avoiding fluctuations due to small sample size effects. This line of reasoning leads to an optimization problem in which a modified version of Expected Shortfall is minimized \cite{SK}. The change that this introduces is known under the name of regularization, and if one choses the L2-norm as a regularizer, then the resulting method is closely related \cite{SK} to support vector regression \cite{CortesVapnik95, Nu-SVM} and robust statistics \cite{huber}.

\section{Regularization from market illiquidity}
 \label{MarketImpact} 
 
To generate cash, an investor has to liquidate (part of) his portfolio. The set up of the portfolio optimization problem above ignored the fact that this liquidation may have an impact on asset prices. 
Let us consider a situation in which an investor holds a portfolio of $N$ assets  and, at time $t$, liquidates a fraction of such portfolio $\vec{w}_t=(w_{1,t},\ldots,w_{N,t})$, with $w_{i,t}$ representing the position of asset $i$ liquidated at time $t$. 
In the following we assume that the liquidation of the portfolio $\vec w_t$ affects prices in a linear way
\begin{equation}
\label{linearw}
\vec p_{t+1}=\vec p_t+\vec x_t-\eta \vec w_t.
\end{equation}
Here $\vec x_t$ is the vector of returns, and $\eta$ is an impact parameter. Notice that investment is taken to move prices in the direction opposite to trading:  selling ($w_{i,t}>0$) will cause prices to fall and buying ($w_{i,t}<0$) will push prices up. The cash flow generated on day $t$ is then given by
\begin{equation}
\label{cflow}
c_t=\vec w_t\cdot \vec p_{t+1}=\vec w_t\cdot \vec p_{t}+\vec w_t\cdot \left(\vec x_t -\eta\vec w_t\right)
\end{equation}
The first part $\vec w_t\cdot \vec p_{t}$ is known at time $t$, so risk only enters in the second part, $\vec w \cdot \vec x -\eta \| \vec w \|^2$, where we have dropped the subscript $t$ to simplify the notation. Similarly to what is done in classical portfolio theory \cite{Markowitz52}, we consider the problem of finding the portfolio of minimal risk, for a given present value $\sum_i w_i p_{i,t}=wN$ of the realized cash flow. 
The parameter $w$ plays the role of a normalization, and is customarily set to one, because risk is usually linear in the size of the portfolio. Here, however, the size of the portfolio matters as the impact of liquidation strategies on prices depends on the size. We therefore keep $w$ as an independent parameter.  
In order to further simplify the notation, we consider $p_{i}=1,~\forall i$, so that we have the constraint
\begin{equation}
\label{ }
\sum_i w_i = wN.
\end{equation}
As before, we take the expected shortfall as a risk measure. The loss is now given by $l(\{w_i\}|\{x_i\})=  - \vec w \cdot \vec x + \eta \| \vec w \|^2$, and we then have to find the minimum of the cost function 
\begin{equation}
\label{LPeta}
E_\eta[v,\{u_{\tau}\}]=(1-\beta) Tv+\sum_{\tau=1}^T u_{\tau}
\end{equation}
under the constraints
\begin{eqnarray}
u_{\tau} & \ge & 0~~\forall\tau, \\
u_{\tau}+v+\sum_{i=1}^N w_i x_{i,\tau} - \eta \| \vec w \|^2& \ge & 0~~\forall\tau, \label{consteta}\\
\sum_i w_i & = & wN.
\end{eqnarray}
All of the $T$ inequality constraints contain a term that is independent of $\tau$ $(\tau=1,\dots, T)$, given by
\begin{equation}
\label{ }
\epsilon= v-\eta \| \vec w \|^2.
\end{equation}
Substitution of $\epsilon + \eta \| \vec w \|^2$ for $v$ in the cost function, Eq. (\ref{LPeta}), and multiplication by ${1 \over 2 (1-\beta) T \eta}$ leads us to the regularized expected shortfall problem proposed recently (see \cite{SK} Eqs. (7)-(10)):
\begin{eqnarray}
&&\min_{\w, {\vec u}, \epsilon}  \left[ {1 \over 2}  \|\w \|^2 +  C \left({1 \over T} \sum_{\tau=1}^{T} u_\tau + \left( 1-\beta \right) \epsilon \right) \right]
\label{newPO} \\
&{\rm s.t.}\;\;\; & \wxk + \epsilon + u_\tau \geq 0; \;\;\; u_\tau \geq 0; \;\;\; \forall \tau, \label{constraints} \\
&& \sum_i w_i = wN. \label{b}
\end{eqnarray}
with
\begin{equation}
\label{Ceta}
C=\frac{1}{2(1-\beta)\eta}.
\end{equation}
We recognize that the term proportional to $\eta$ in Eq. (\ref{cflow}) acts as a regularizer. 

To develop a simple intuitive argument, imagine that there are two portfolios $\vec{w}^+$ and $\vec{w}^-$, each properly normalized (i.e. $\sum_i w_i^\pm =  wN$), with $\vec{w}^+\vec{x}_\tau\ge \vec{w}^-\vec{x}_\tau$ for all $\tau=1,\ldots,t$ and $\vec{w}^+\vec{x}_\tau > \vec{w}^-\vec{x}_\tau$ for at least one $\tau$. Then, when $\eta=0$, minimal Expected Shortfall would be realized by selling $K$ units of $\vec{w}^-$ and buying $K+1$ units of $\vec{w}^+$, with $K\to \infty$. This, as shown in \cite{kondor2008}, is the origin of the instability in coherent risk measures. Such infinite returns cannot be realized, however, by liquidating a real portfolio because prices will adjust. In the linear approximation discussed here, when $\eta>0$, the investment behavior discussed above is going to modify future returns, because $x_{i,t+1}\to x_{i,t+1}-\eta w_{i}$, 
thereby eliminating the apparent arbitrage. This effect reflects precisely the logic behind the no-arbitrage hypothesis.

\section{Behavior of large random minimal risk portfolios under regularized expected shortfall} 
 \label{regularize} 
We have argued that the observed instability can be alleviated by regularization \cite{SK}. In particular, regularized portfolio selection under Expected Shortfall is related to support vector regression \cite{SK}. This opens up the way to apply the existing support vector algorithms, duly modified, to portfolio selection. 

We will show here that, if we make an assumption on the distribution of the underlying data, then we can make progress by analytical means as well. We follow the calculation in \cite{CKM07} to see how the regularizer takes care of the instability. In the course of the calculation we make use of the powerful techniques of the statistical physics of random systems. Some details of the derivation are reported in the Appendix, but these details are not necessary for the understanding of the result which is a very plausible extension of that in \cite{CKM07}. Therefore, here we just quote the result and discuss its consequences, namely the removal of the singularity of the risk measure. We also refer the interested reader to the related literature within statistical learning theory, such as \cite{Opar950,OpHa95d0,DiOpSo990,MaOp05a0} and references therein.

In the previous sections both statistical and financial considerations led us to a cost function of the form 
$$E[v,\{u_{\tau}\}]=(1-\beta) T\epsilon+\sum_{\tau=1}^T u_{\tau}+\tilde{\eta } \|w\|^2,$$
where $\tilde{\eta}$ can be expressed in terms of $C$ or $\eta$.
Starting from this expression, and averaging over the returns $\{x_{i,\tau}\}$ drawn from a Gaussian distribution\footnote{Here $x_{i,t}$ are taken as i.i.d. Gaussian variables with zero mean and variance $1/\sqrt{N}$. The latter ensures a meaningful limit $N,T\to\infty$ with $N/T=n$ constant, and is also realistic for typical cases where $N\sim 10^3-10^4$.} via the method described in the Appendix, one arrives at a generalized cost function given in terms of three variational parameters  $\Delta$, 
$q_0=\tilde{q}_0\Delta^2=\sum w_i^2/N$, and $\epsilon=\tilde{\epsilon}\Delta$, as in \cite{CKM07}
\begin{eqnarray}
E(\tilde{\epsilon},\tilde{q}_0,\Delta)&=&
\frac{w^2}{2\Delta}+\Delta\left[t(1-\beta)\tilde{\epsilon}-\frac{\tilde{q}_0}{2}+\frac{t}{2\sqrt{\pi}}\int_{-\infty}^{\infty}ds e^{-s^2}
g(\tilde{\epsilon}+s \sqrt{2 \tilde{q}_0})\right]+\tilde{\eta} \tilde{q}_0\Delta^2\qquad,
\label{ERPO}
\end{eqnarray}
where 
\begin{equation}
    g(x)= \left\{ \begin{array}{cc} 0 ,&    x\ge 0\\
    x^2 , &  -1\le x\le 0\\
    -2 x-1, & x<-1
     \end{array} \right. .
\end{equation}
The difference with respect to \cite{CKM07} is that now we have an additional term proportional to $q_0$. Indeed, if one considers the definition of $\tilde q_0$, the extra term $\tilde{\eta}\tilde{q}_0\Delta^2$ precisely maps into the term $\tilde{\eta} \|w\|^2$ added to the objective function. 

Let us now discuss how the term proportional to $\|w\|^2$ in the cost function prevents the instability in the portfolio optimization problem.
The first order conditions on \eqref{ERPO} with respect to the three variational parameters read
\begin{equation}\label{q2}
-1+\frac{t}{\sqrt{2\pi \tilde{q}_0}}\int ds e^{-s^2} s g'(\tilde{\epsilon}+s \sqrt{2 \tilde{q}_0})+2 \tilde{\eta}\Delta=0,
\end{equation}
\begin{equation}\label{v2}
1-\beta+\frac{1}{2\sqrt{\pi}}\int ds e^{-s^2} g'(\tilde{\epsilon}+s \sqrt{2 \tilde{q}_0})=0,
\end{equation}

\begin{equation}\label {delta2}
-\frac{w^2}{2\Delta^2}+t(1-\beta)\tilde{\epsilon}-\frac{\tilde{q}_0}{2}+\frac{t}{2\sqrt{\pi}}\int ds e^{-s^2}  g(\tilde{\epsilon}+s \sqrt{2 \tilde{q}_0})+2\tilde{\eta} \Delta \tilde{q}_0= 0.
\end{equation}

\begin{center}
\begin{figure}[h]
\begin{center}
\includegraphics[width=8cm]{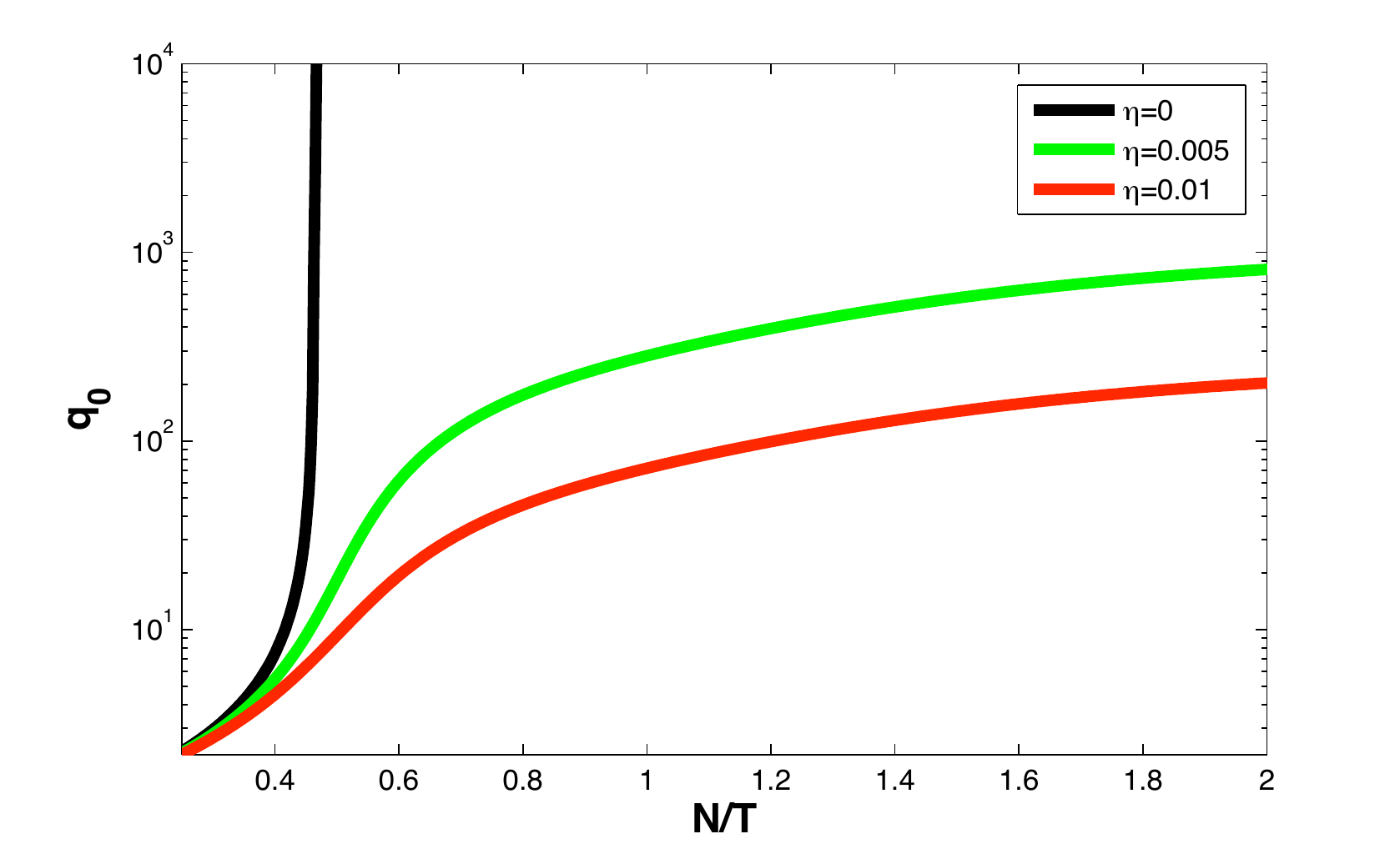}
\caption{\footnotesize {\textit{$q_0$ as a function of $N/T$ for different values of $\tilde{\eta}$ and $\beta=0.7$.}}} \label{q0}
\end{center}
\end{figure}
\end{center}

\begin{center}
\begin{figure}[h]
\begin{center}
\includegraphics[width=8cm]{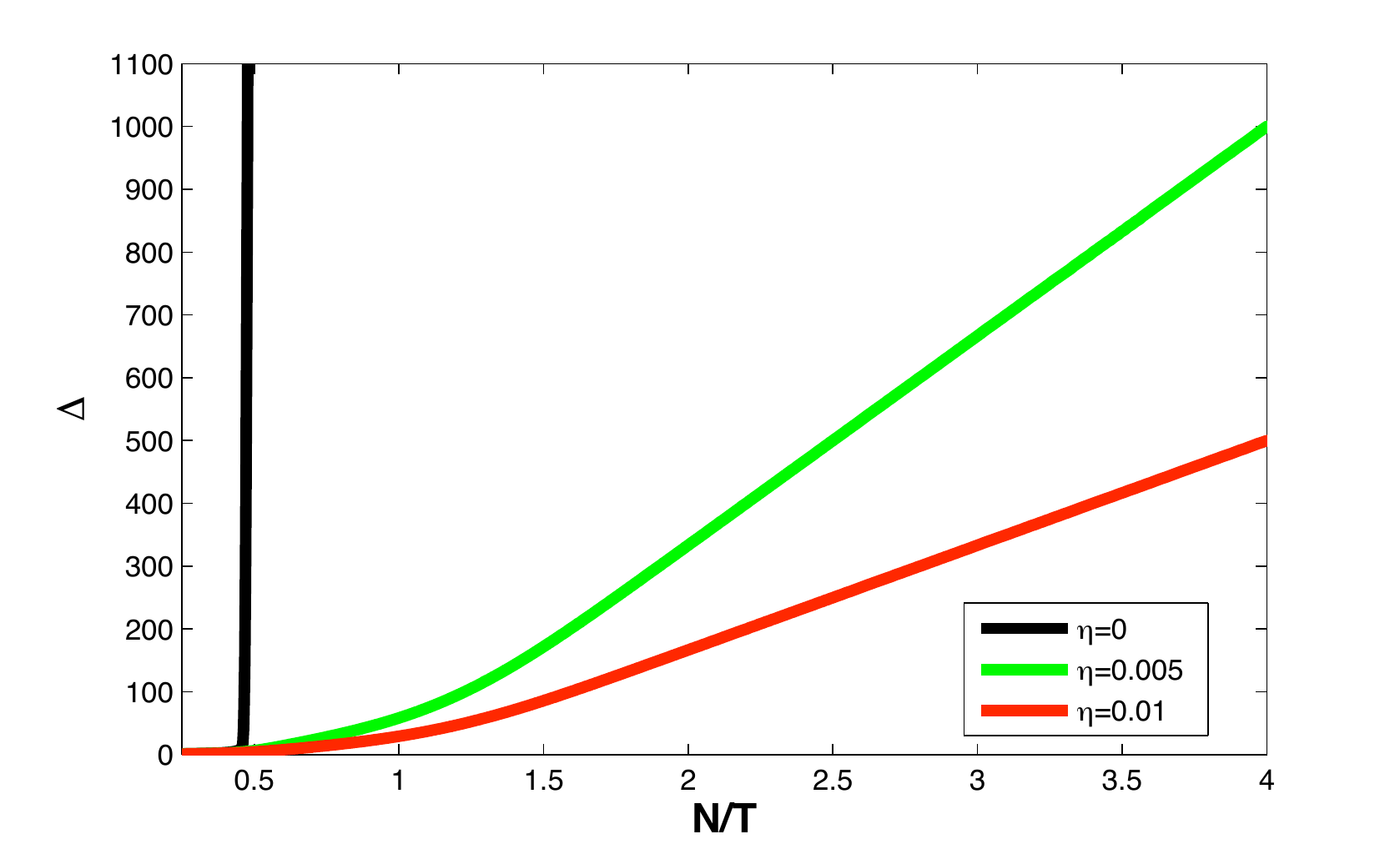}
\caption{\footnotesize {\textit{The susceptibility as a function of $N/T$ for different values of $\tilde{\eta}$ and $\beta=0.7$.}}} \label{susc}
\end{center}
\end{figure}
\end{center}

\begin{center}
\begin{figure}[h]
\begin{center}
\includegraphics[width=8cm]{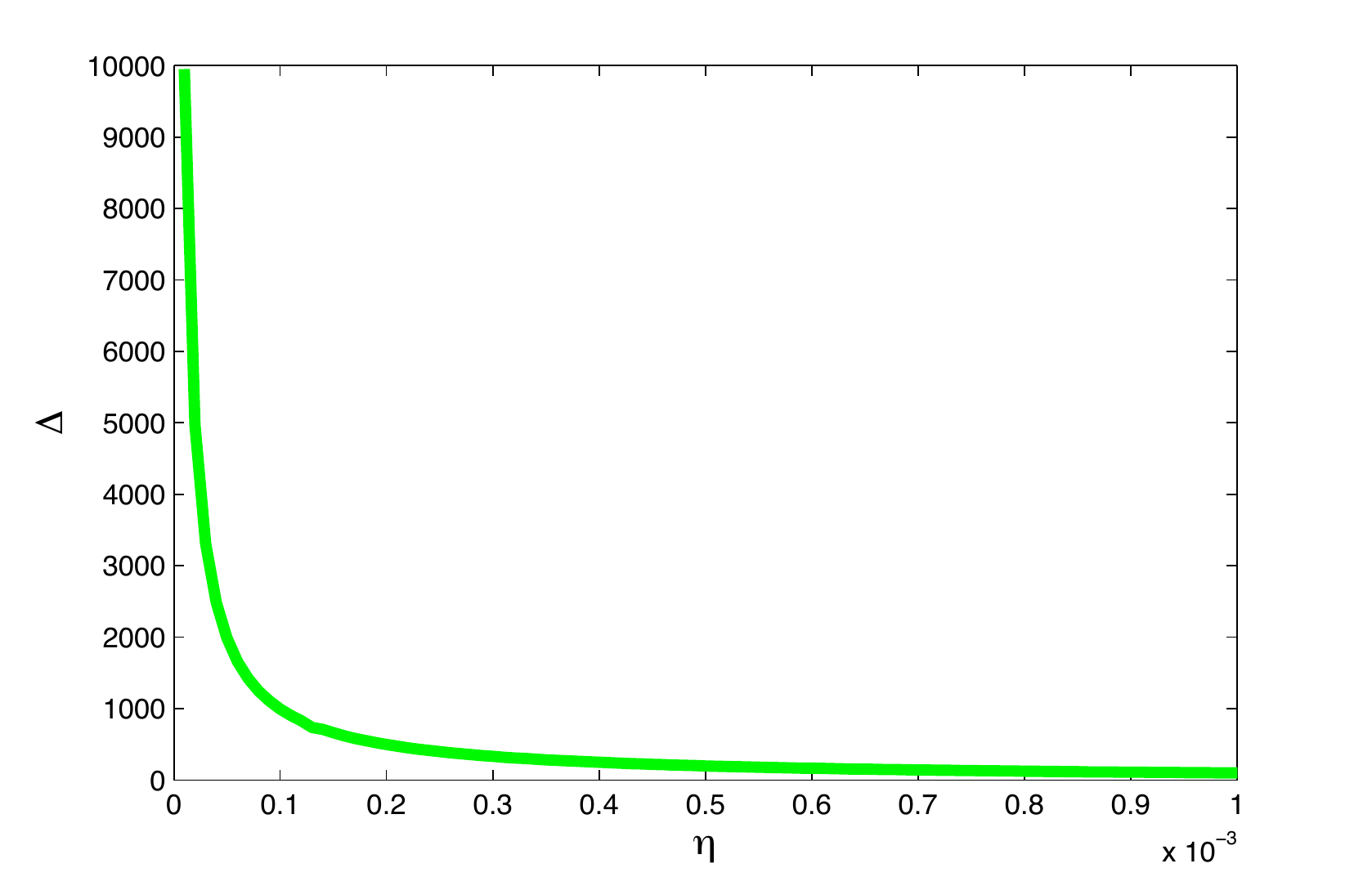}
\caption{\footnotesize {\textit{The susceptibility as a function of $\tilde{\eta}$ for the case $t=1.5$ and $\beta=0.7$ .}}} \label{susc_eta}
\end{center}
\end{figure}
\end{center}

\noindent In the original, non-regularized problem in \cite{CKM07} the instability of Expected Shortfall was indicated by the divergence of the parameter $\Delta$ which we will, therefore, call the susceptibility here. The divergence of this susceptibility is thus the signature of the instability, which made it expedient to rescale the variables as $\tilde{\epsilon}=\epsilon/\Delta$ and 
$\tilde{q}_0=q_0/\Delta^2$ in the first order conditions. Notice that the variables $\tilde{\epsilon}$ and $\tilde{q}_0$ are already finite \footnote{The divergence of $\tilde{\epsilon}$ and  $\tilde{q}_0$ is prevented by equation \eqref{v2}, that admits solutions only if the two variables are finite.}.
This should be sufficient to conclude that  all integrals over the variable $s$ are finite.
In order to see if a solution with divergent susceptibility can exist, we now let $\Delta\to\infty$ in the first order conditions.
We first note that, in order for \eqref{delta2} to be satisfied, $\tilde{q}_0\Delta$ has to remain finite as $\Delta\to\infty$, i.e. $\tilde{q}_0=\alpha/\Delta$ with $\alpha$ finite.
However, this is in contrast with a similar constraint we can deduce from equation \eqref{q2},
where we find that a solution exists only if $\Delta\sqrt{\tilde{q}_0}$ is finite.
Indeed if we  multiply all terms of \eqref{q2} for $\sqrt{\tilde{q}_0}$ and impose $\tilde{q}_0=\alpha/\Delta$ we see that all the terms are bounded except the last one which diverges as $\sqrt{\Delta}$. 
We thus conclude that no solution with divergent susceptibility can be found as long as $\tilde{\eta}>0$.\\
The numerical solution of the first order conditions confirms this prediction. 
In figures \ref{q0} and \ref{susc} we show the behavior of $q_0=\frac{1}{N}\sum_i w_i^2$ and of $\Delta$. We can clearly observe that the divergence, which is present for $\tilde{\eta}=0$, disappears as soon as $\tilde{\eta}>0$.  This is further confirmed  by figure \ref{susc_eta}, where we show that, in the unfeasible region of the original problem, $\Delta$ diverges at $\tilde{\eta}=0$.

\section{Discussion} 

Let us now comment on the generality of the result. Concerning the linear assumption in Eq. (\ref{linearw}), which is standard in the econometric literature \cite{hasbrouk1991}, we observe that the estimate of market impact functions is a matter of active current research. Most of the empirical evidence suggests a convex shape \cite{Eisler_etal2009}, with the price impact growing slower than $w$.
In double auction markets, if one restricts attention to the instantaneous impact of market orders, the effect on the price depends on the shape of the order book. In order to discuss this case in some more detail, let $\rho_i(p,t)$ be the density of limit orders for asset $i$ at time $t$, and consider the situation where a market order for a quantity $w_i$ arrives at time $t$. If $p_{i,t-1}$ is the current price and $p_{i,t-1}+x_{i,t}$ is the price (of the transaction which occurred) just before the order arrives, then the price $p_{i,t}$ at which the transaction will take place is given by  
\begin{equation}
\label{orderbook}
w_i=\int_{p_{i,t-1}+x_{i,t}}^{p_{i,t}}\! dp \rho_i(p,t).
\end{equation}
A linear impact, as the one assumed in Eq. (\ref{linearw}), then corresponds to an order book with a constant density of limit orders. 
Hence, a measure of $\eta$ is given by the density of the order book close to the best bid/ask. 
Since the density of the order book fluctuates and liquidity varies across assets, $\eta_{i,t}$ could also be taken as an asset dependent stochastic quantity. 

Then the computation of the ES can still be performed in terms of the cost function (\ref{LPeta}), but now with the $T$ constraints
\begin{equation}
u_{\tau}+v+\sum_{i=1}^N w_i x_{i,\tau} - \sum_i \eta_{i,\tau} w_i^2 \ge  0 \label{consteta_gen}
\end{equation}
in place of Eq. (\ref{consteta}). 
In this case the mapping to a simple $L2$ regularizer that we have laid out in section \ref{MarketImpact}, is then complicated by the fact that the impact term depends on $\tau$ and cannot be absorbed into a $\tau$ independent constant.
Nevertheless we do not expect the essential features of the problem to change with respect to the case of a constant $\eta$.

Note furthermore that different assumptions for the market impact function lead to different regularizers. For example, considering the instantaneous impact and Eq. (\ref{orderbook}), in the presence of a bid-ask spread, we expect the price to bounce from the bid to the ask, depending on the direction of trading (i.e. on the sign of $w_i$). This suggests a term proportional to the sign of $w_i$ in the equation for the price, which, in turn, would then introduce an $L1$ regularizer. The choice of the regularizer and the behavior of regularized portfolio optimization under the different choices is a rich subject which deserves a separate treatment. However, this choice is related to market impact and liquidity risk considerations.

Notice, finally, that the Maximal Loss limit, $\beta\to 1$, is non-trivial. In this limit, the Expected Shortfall reduces to the Maximal Loss (ML), which reads
\begin{equation}
\label{ML}
ML(\mathbf{w})=\max_{t=1,\ldots,T} \left[-\sum_i w_i x_{i,t}\right].
\end{equation}
If we include the market impact term, i.e. if $x_{i,t}\to x_{i,t}-\eta w_i$ then 
\[
ML(\mathbf{w})=\eta \sum_i w_i^2+\max_{t=1,\ldots,T} \left[-\sum_i w_i x_{i,t}\right],
\]
which is clearly finite, for finite $\eta$. However, if the $\beta\to 1$ limit was taken in Eq. (\ref{Ceta}), one would finds that $C\to\infty$, suggesting that regularization would disappear. The correct limit is recovered (see Appendix) by keeping $T(1-\beta)=1$, i.e. by taking $\beta = 1-1/T$, and then eventually letting $T \to \infty$. Therefore, $C=T/2\eta$, which implies that for large $T$ and small $\eta$, liquidity considerations provide only a weak regularizer in the limit of Maximal Loss.

\section{Conclusion} 
We have shown that considering an optimal liquidation policy for a portfolio automatically leads to regularized portfolio optimization. Liquidity considerations provide a way to choose the regularizer. When the market impact is assumed to be linear, we obtain the L2-norm as a regularizer. The typical behavior of large portfolios changes due to the regularization -- the otherwise divergent fluctuations in the optimal solution are now tamed. This indicates that regularized portfolio optimization is a better investment strategy for large portfolios than traditional portfolio optimization, which minimizes only the empirical risk. 

\section{Acknowledgments}

I.K.  has been supported by the "Cooperative Center for Communication Networks Data
Analysis", sponsored by the Teller program of the National Office of  Research and Technology
under grant No. KCKHA005. He is also grateful for the hospitality extended to him at the University of Hawaii at Manoa, Honolulu, during the preparation of this work.

\appendix
\section{The replica calculation}

We present here the calculation we use to solve the following optimization problem:
find the minimum of the cost function
$$E[\epsilon,\{u_{\tau}\}]=(1-\beta) T\epsilon+\sum_{\tau=1}^T u_{\tau}+\tilde{\eta } \|w\|^2$$
under the constraints
$$u_{\tau}\ge0,$$ $$u_{\tau}+\epsilon+\sum_{i=1}^N x_{i,\tau} w_i\ge 0$$ and $$\sum_i w_i =wN.$$

The underlying process governing the returns is assumed to be i.i.d normal.
We are using the machinery of statistical physics of random systems, and the terminology will be chosen accordingly. The calculation begins with regarding the cost function as a kind of abstract "Hamiltonian" or energy functional and introducing a fictitious temperature associated with the system. The reason for this is purely technical, and the original problem will be recovered in the zero temperature limit.

Given a realization for the history of returns $\{x_{i,\tau}\}$, the calculation proceeds by considering the partition function or generating functional
\be
Z_{\gamma}(\{x_{i,\tau}\})=\int_{V(\{x_{i,\tau}\})} d \vec{Y} e^{-\gamma E[\vec{Y}]},
\ee
where $\gamma$ is the inverse temperature, and we have used the notation $\vec{Y}$ to indicate the set of variables, and $V(\{x_{i,\tau}\})$ represents the portion of phase space where all constraints are satisfied. The minimum cost can then be computed as
\be
\lim_{N\to\infty}\lim_{\gamma\to\infty} -\frac{\log Z_{\gamma}(\{x_{i,\tau}\})}{N\gamma}
\ee
In order to compute typical properties of the ensemble, we average over the probability distribution of returns, that is we compute the average of $\log Z_{\gamma}(\{x_{i,\tau}\})$.
This can be achieved through the replica trick by exploiting the identity
\be
\langle \log Z\rangle=\lim_{n\to 0}\frac{\partial Z^n}{\partial n}.
\ee

The replicated partition function, corresponding to the partition function of $n$ copies of the system
can be computed as \footnote{In the following calculation we don't keep track of multiplicative factors which are constant and do not contribute to the final free energy.}
\begin{eqnarray}
Z_{\gamma}^n[{x_{i,\tau}}] &=& \int_{-\infty}^{\infty} \prod_{a=1}^n d\epsilon^a
\int_0^{\infty}\prod_{\tau=1}^T\prod_{a=1}^n du_{\tau}^a\int_{-\infty}^{\infty} \prod_{i=1}^N \prod_{a=1}^n d
w_i^a\int_{-\infty}^{\infty}\prod_{a=1}^n d\lambda^a \\
 &\times& \int_{0}^{\infty} \prod_{\tau=1}^T\prod_{a=1}^n d
\mu_{\tau}^a \int_{-\infty}^{\infty}\prod_{\tau=1}^T \prod_{a=1}^n d\hat{\mu}_{\tau} \exp\left\{\sum_a \lambda^a
(\sum_i w_i^a-wN)\right\}\\
&\times&\prod_{\tau}\exp\left\{\sum_a i \hat{\mu}_{\tau}^a\left(u_{\tau}^a+\epsilon^a+\sum_i x_{i,\tau}w_i^a-\mu_{\tau}^a \right)\right\}\\
&\times& \exp\left\{-\gamma\sum_a(1-\beta)T \epsilon^a-\gamma\sum_{a,\tau} u_{\tau}^a-\gamma\tilde{\eta}\sum_i
{w_i^a}^2\right\}.
\end{eqnarray}
Averaging over the quenched variables $\{x_{i,\tau}\}$ and introducing the overlap matrix
$Q_{a,b}=\frac{1}{N}\sum_i w_i^a w_i^b$ one obtains\footnote{The variables $x_{i,\tau}$ are assumed to have zero average and variance $1/N$. Therefore, to leading order in $N$,
\begin{eqnarray}
\left\langle e^{x_{i,\tau} \Gamma_{i,\tau}}\right\rangle & \simeq & \left\langle 1+ x_{i,\tau} \Gamma_{i,\tau}+\frac{1}{2}\left(x_{i,\tau} \Gamma_{i,\tau}\right)^2+\ldots\right\rangle \\
 & = & 1+\frac{1}{2N}\Gamma_{i,\tau}^2+\ldots\\
 & = & e^{ \frac{1}{2N}\Gamma_{i,\tau}^2}(1+\ldots)
\end{eqnarray}
where $\Gamma_{i,\tau}=i\sum_a\hat\mu_\tau^a w_i^a$ and $\ldots$ stands for higher order terms in an expansion in powers of $1/N$. In other words, as long as the variance of $x_{i,\tau}$ is well defined, the results carry through irrespective of the specific distribution of $x_{i,\tau}$ (i.e. no assumption of Gaussian returns is needed). The calculation is performed for the simple case of independent returns. The case of returns with a given correlation matrix can also be dealt with.}.
\begin{eqnarray*}
Z_{\gamma}^n[{x_{i,\tau}}] &=& \int[D\epsilon][Du][Dw][D\lambda][D\mu][D\hat{\mu}][DQ][D\hat{Q}] \exp\left\{\sum_a
\lambda^a(\sum_i w_i^a-wN)\right\}\\
&\times& \exp\left\{-\gamma\sum_a(1-\beta)T \epsilon^a-\gamma\sum_{a,\tau}
u_{\tau}^a-\gamma\tilde{\eta}\sum_i
{w_i^a}^2\right\}\prod_{\tau}\exp\left\{-\frac{1}{2}\sum_{a,b}\hat{\mu}_{\tau}^aQ_{a,b}\hat{\mu}_{\tau}^b\right\}\\
&\times& \exp\left\{\sum_{a,b}\hat{Q}_{a,b}\left(N Q_{a,b}-\sum_i w_i^a w_i^b\right)\right\}\\
&\times&\prod_{\tau}\exp\left\{i\sum_a\hat{\mu}_{\tau}^a\left(u_{\tau}^a+\epsilon^a-\mu_{\tau}^a \right)\right\}.
\end{eqnarray*}
We can now perform the Gaussian integral over the variables $\{\hat{\mu}_{\tau}^a\}$:
\begin{eqnarray*}
Z_{\gamma}^n[{x_{i,\tau}}] &=& \int[D\epsilon][Du][Dw][D\lambda][D\mu][DQ][D\hat{Q}] \exp\left\{\sum_a
\lambda^a(\sum_i w_i^a-wN)\right\}\\
&\times& \exp\left\{-\gamma\sum_a(1-\beta)T \epsilon^a-\gamma\sum_{a,\tau}
u_{\tau}^a-\gamma\tilde{\eta} \sum_i
{w_i^a}^2\right\}\exp\left\{\sum_{a,b}\hat{Q}_{a,b}\left(N Q_{a,b}-\sum_i w_i^a w_i^b\right)\right\}\\
&\times& \prod_{\tau}\exp\left\{-\frac{1}{2}\sum_{a,b}\left(u_{\tau}^a+\epsilon^a-\mu_{\tau}^a\right)Q_{a,b}^{-1}\left(u_{\tau}^b+\epsilon^b-\mu_{\tau}^b\right)\right\}\\
&\times& \exp\left\{-\frac{T}{2}{\rm tr}\log Q\right\}.
\end{eqnarray*}
We are now allowed to perform a Gaussian integration over the variables $\{w_i^a\}$, which is going to bring
the inverse of the operator $\hat{Q}_{a,b}+\gamma\tilde{\eta}\delta_{a,b}$ into the game:
\begin{eqnarray*}
Z_{\gamma}^n[{x_{i,\tau}}] &=& \int[D\epsilon][Du][D\lambda][D\mu][DQ][D\hat{Q}] \exp\left\{\sum_a
-w\lambda^aN\right\}\\
&\times& \exp\left\{-\gamma\sum_a(1-\beta)T \epsilon^a-\gamma\sum_{a,\tau} u_{\tau}^a\right\}\\
&\times& \prod_{\tau}\exp\left\{-\frac{1}{2}\sum_{a,b}\left(u_{\tau}^a+\epsilon^a-\mu_{\tau}^a\right)Q_{a,b}^{-1}\left(u_{\tau}^b+\epsilon^b-\mu_{\tau}^b\right)\right\}\\
&\times& \exp\left\{\sum_{a,b}\hat{Q}_{a,b}NQ_{a,b}\right\}
\exp\left\{-\frac{T}{2}{\rm tr}\log Q\right\}\exp\left\{-\frac{nN}{2}\log 2\right\}\\
&\times& \exp\left\{-\frac{N}{2}{\rm tr}\log
(\hat{Q}+\gamma\tilde{\eta}\delta_{a,b})\right\}\exp\left\{\frac{N}{4}\sum_{a,b}\lambda^a\left(\hat{Q}_{a,b}+\gamma\tilde{\eta}\delta_{a,b}\right)^{-1}\lambda^b\right\}.
\end{eqnarray*}
Integrating now over the $\{\lambda^a\}$ we obtain
\begin{eqnarray*}
Z_{\gamma}^n[{x_{i,\tau}}] &=& \int[D\epsilon[Du][D\mu][DQ][D\hat{Q}]
 \exp\left\{-\gamma\sum_a(1-\beta)T \epsilon^a-\gamma\sum_{a,\tau} u_{\tau}^a\right\}\\
&\times& \prod_{\tau}\exp\left\{-\frac{1}{2}\sum_{a,b}\left(u_{\tau}^a+\epsilon^a-\mu_{\tau}^a\right)Q_{a,b}^{-1}\left(u_{\tau}^b+\epsilon^b-\mu_{\tau}^b\right)\right\}\\
&\times& \exp\left\{N\sum_{a,b}\hat{Q}_{a,b}Q_{a,b}\right\}
\exp\left\{-\frac{T}{2}{\rm tr}\log Q\right\}\exp\left\{-\frac{nN}{2}\log 2\right\}\\
&\times& \exp\left\{-\frac{N}{2}{\rm tr}\log
(\hat{Q}+\gamma\tilde{\eta}\delta_{a,b})\right\}\exp\left\{-Nw^2\sum_{a,b}(\hat{Q}_{a,b}+\gamma\tilde{\eta}\delta_{a,b})\right\}.
\end{eqnarray*}
Introducing the variables $y_{\tau}^a=\mu_{\tau}^a-u_{\tau}^b$ and
$z_{\tau}^a=\mu_{\tau}^a+u_{\tau}^b$ and integrating over the $\{z_{\tau}^a\}$ one is left with
\begin{eqnarray*}
Z_{\gamma}^n[{x_{i,\tau}}] &=& \int[D\epsilon][DQ][D\hat{Q}]\\
&\times& \exp\left\{-Nw^2\sum_{a,b}(\hat{Q}_{a,b}+\gamma\tilde{\eta}\delta_{a,b})-\gamma(1-\beta)\sum_a T
\epsilon^a+N\sum_{a,b}\hat{Q}_{a,b}Q_{a,b}\right\}\\
 &\times&\exp\left\{-Tn \log\gamma -\frac{T}{2}{\rm tr}\log Q-\frac{N}{2}{\rm tr}\log(\hat{Q}+\gamma\tilde{\eta}\delta_{a,b})-\frac{nN}{2}\log 2\right\}\\
 &\times&\exp\left\{ T\log Z_{\gamma}(\{\epsilon^a,Q\})\right\},
\end{eqnarray*}
where we have defined
\begin{eqnarray*}
Z_{\gamma}(\{\epsilon^a,Q\})&=&\int \prod_a dy^a \exp\left\{-\frac{1}{2}\sum_{a,b}(y^a-\epsilon^a)Q_{a,b}^{-1}(y^b-\epsilon^b)\right\}\\
&\times& \exp\left\{\gamma\sum_a y^a\theta(-y^a)\right\}.
\end{eqnarray*}
We now take the replica symmetric (RS) ansatz
\begin{equation}\label{RS}
    Q_{a,b}= \left\{ \begin{array}{cc} q_1 ,&    a =b\\
    q_0 , &  a\neq b \end{array} \right.
\end{equation}
\begin{equation}\label{RS1}
    \hat{Q}_{a,b}= \left\{ \begin{array}{cc} \hat{q}_1 ,&    a =b  \\
    \hat{q}_0 , &  a\neq b . \end{array} \right.
\end{equation}
and we define the susceptibility $\Delta q=q_1-q_0$ as well as $\Delta\hat{q}=\hat{q}_1-\hat{q}_0$. 
For $Q^{-1}$ we then have
\begin{equation}\label{RS2}
    Q_{a,b}^{-1}= \left\{ \begin{array}{cc} (\Delta q-q_0)/(\Delta q)^2+ \mathcal{O}(n) ,&    a =b\\
    -q_0/(\Delta q)^2+ \mathcal{O}(n), &  a\neq b \end{array} \right.
\end{equation}
The effective partition function $Z_{\gamma}(\{\epsilon^a,Q\})$ reads
\begin{eqnarray*}
Z_{\gamma}(\{\epsilon^a,R^a,Q\})&=&\int \prod_a dx^a \exp\left\{-\frac{1}{2}\sum_{a,b}x^a Q_{a,b}^{-1}x^b\right\}\\
&\times& \exp\left\{\gamma\sum_a (x^a+\epsilon^a)\theta(-x^a-\epsilon^a)\right\},
\end{eqnarray*}
where we have defined $x^a=y^a-\epsilon^a$. By introducing a Gaussian variable $s$ with measure
$dP_{q_0}(s)=\frac{ds}{\sqrt{2\pi q_0}}e^{-s^2/2 q_0}$ we obtain, in the limit $n\to 0$,
\begin{equation*}
\frac{1}{n}\log(Z_{\gamma}(\{\epsilon^a,Q\})=\frac{q_0}{2\Delta q}+\int dP_{q_0}(s)\log B_{\gamma}(s,\epsilon,\Delta q),
\end{equation*}
where 
\begin{eqnarray*}
B_{\gamma}(s,\epsilon,\Delta q)=\int dx \exp\left\{-\frac{(x-s)^2}{2\Delta q}+\gamma(x+\epsilon)\theta(-x-\epsilon)\right\}.
\end{eqnarray*}

If we also consider that
$${\rm tr}\log Q=n(\log\Delta q+q0/\Delta q)$$ and $${\rm tr}\log (\hat{Q}+\gamma\tilde{\eta}\delta_{a,b})=n(\log(\Delta\hat{q}+\gamma\tilde{\eta})+\hat{q_0}/(\Delta
\hat{q}+\gamma\tilde{\eta})),$$ we finally obtain the free energy
\begin{eqnarray*}
-\frac{\gamma F(q_0,\Delta q, \hat{q_0},\Delta\hat{q},\epsilon)}{nN}&=& q_0\Delta\hat{q}+\hat{q}_0\Delta
q+\Delta
q\Delta\hat{q}-w^2\left(\Delta\hat{q}+\gamma\tilde{\eta}\right)-\gamma t(1-\beta) \epsilon\\
&-&t\log \gamma +t\int dP_{q_0}(s) \log B_{\gamma}(\epsilon,s,R,\Delta q)-\frac{t}{2}\log\Delta q\\
&-& \frac{\log
2}{2}-\frac{1}{2}\left(\log(\Delta\hat{q}+\gamma\tilde{\eta})+\frac{\hat{q}_0}{\Delta\hat{q}+\gamma\tilde{\eta}}\right),
\end{eqnarray*}
where we have put $T=tN$. From the saddle point equations for $\hat{q}_0$ and $\Delta\hat{q}$ we get
\begin{eqnarray*}
\Delta\hat{q}+\gamma\tilde{\eta}&=&\frac{1}{2\Delta q}\\
\hat{q}_0&=&\frac{w^2-q_0}{2(\Delta q)^2}.\\
\end{eqnarray*}
Exploiting these relations the free energy becomes
\begin{eqnarray*}
-\gamma f(\epsilon,q_0,\Delta q)&=&-\frac{\gamma F}{nN}=\frac{1}{2}-t\log\gamma-\gamma t(1-\beta) \epsilon\\
&+& t\int dP_{q_0}(s) \log B_{\gamma}(\epsilon,s,R,\Delta q)+\frac{1-t}{2}\log\Delta q+\frac{q_0-w^2}{2\Delta q}-\gamma\tilde{\eta}q_0.
\end{eqnarray*}

Notice that 
\begin{equation}
\label{ }
\Delta q=\frac{1}{2N}\sum_i (w_i^a-w_i^b)^2
\end{equation}
is the squared distance between two approximate solution of the optimization problem, drawn with a Gibbs measure with energy $E$. As $\gamma\to\infty$ the Gibbs measure gets more and more peaked on the optimal solution. If the latter is unique, we expect $\Delta q\to 0$. Indeed, given that the measure is nearly Gaussian, we expect $\Delta q\sim 1/\gamma$. Hence, in the large $\gamma$ limit, it is natural to rescale $\Delta q=\Delta/\gamma$  keeping $\epsilon$ and $q_0$ independent
of $\gamma$. In this limit we obtain the energy function
\begin{eqnarray*}
E&=&t(1-\beta)\epsilon-\frac{q_0-w^2}{2\Delta}-\int_{-\infty}^{-\Delta} \frac{d x}{\sqrt{2\pi q_0}}e^{-(x-\epsilon)^2/(2q_0)}\left(x+\frac{\Delta}{2}\right)\\
&+&\frac{t}{2\Delta}\int_{-\Delta}^0\frac{d x}{\sqrt{2\pi q_0}}e^{-(x-\epsilon)^2/(2q_0)}x^2.
\end{eqnarray*}

We now define $\tilde{x}= x/\Delta$, $\tilde{\epsilon}=\epsilon/\Delta$ and $\tilde{q}_0=q_0/\Delta^2$. The reason for this change of variables is that we want to expose the singular behavior at the phase transition in terms of a single divergent quantity\footnote{It helps to note that $\Delta$ is the susceptibility.} $\Delta$. Hence, we anticipate that $\tilde{\epsilon}$ and $\tilde{q}_0$ are going to attain finite values at the transition. In terms of the rescaled variables, we have
\begin{eqnarray*}
E(\tilde{\epsilon},\tilde{q}_0,\Delta)&=& \label{tildecost}
\frac{w^2}{2\Delta}+\Delta\left[t(1-\beta)\tilde{\epsilon}-\frac{\tilde{q}_0}{2}+\frac{t}{2\sqrt{\pi}}\int_{-\infty}^{\infty}ds e^{-s^2}
g(\tilde{\epsilon}+s \sqrt{2 \tilde{q}_0})\right]+\tilde{\eta}t\tilde{q_0}\Delta^2
\end{eqnarray*}
where
\begin{equation}
    g(x)= \left\{ \begin{array}{cc} 0 ,&    x\ge 0\\
    x^2 , &  -1\le x\le 0\\
    -2 x-1, & x<-1
     \end{array} \right.
\end{equation}
and $\tilde{q}_0$ and $\tilde{\epsilon}$ are the solutions of the saddle point equations
\begin{equation}\label{q}
-1+\frac{t}{\sqrt{2\pi \tilde{q}_0}}\int ds e^{-s^2} s g'(\tilde{\epsilon}+s \sqrt{2 \tilde{q}_0})+2 \tilde{\eta}\Delta=0,
\end{equation}
\begin{equation}\label{v}
1-\beta+\frac{1}{2\sqrt{\pi}}\int ds e^{-s^2} g'(\tilde{\epsilon}+s \sqrt{2 \tilde{q}_0})=0,
\end{equation}

\begin{equation}\label {delta}
-\frac{w^2}{2\Delta^2}+t(1-\beta)\tilde{\epsilon}-\frac{\tilde{q}_0}{2}+\frac{t}{2\sqrt{\pi}}\int ds e^{-s^2}  g(\tilde{\epsilon}+s \sqrt{2 \tilde{q}_0})+2\tilde{\eta} \Delta \tilde{q}_0= 0.
\end{equation}

\section{The Maximal Loss Problem}
We show here how to recover the correct $\beta\to 1$ limit leading to the Maximal Loss problem.
The problem of finding the set of weights that minimizes the Maximal Loss \eqref{ML} can be cast into that of finding the minimum of the cost function
\begin{equation}
 E[u]=u \label{cost1}
 \end{equation}
 under the constraints 
$$u+\sum_i w_{i}x_{i,t}\ge 0~~~\forall t$$ and $$\sum_i w_i=wN.$$
Let us show how this can be recovered starting from the general problem for the Expected Shortfall 
\begin{eqnarray}
\label{cost2}E[\epsilon,\{u_{\tau}\}]&=&(1-\beta) T\epsilon+\sum_{\tau=1}^T u_{\tau},\\
\label{cos1} u_{\tau}&\ge&0~~\forall\tau ,\\
u_{\tau}+\epsilon+\sum_{i=1}^N x_{i,\tau} w_i&\ge& 0~~\forall\tau\label{cos2},\\ 
\sum_i w_i &=&wN.
\end{eqnarray}

The first observation is that, for $\epsilon\ge ML$,  \eqref{cos2} is satisfied for any set of $\{u_{i,\tau}\}$ satisfying \eqref{cos1}. 
The minimum of the cost function can then be obtained by taking $\epsilon$ equal to the Maximal Loss and $u_{i,\tau}=0~\forall i,\tau$. By comparing the resulting  expression for \eqref{cost2} with \eqref{cost1}, we can see that the two are equivalent if we keep $T(1-\beta)=1$.
If we now introduce the regularization, the cost function for the Maximal Loss problem reads 
$$E[u,\{w_i\}]=u+\frac{T}{2 C} \| \vec w \|^2.$$ 
As in section \ref{MarketImpact}, an equivalent expression can be obtained by introducing the effect of the price impact. The two approaches are equivalent  once we have taken
\be
C=\frac{T}{2\eta}.\label{Cetabeta1}
\ee
Notice that one can derive Eq. (\ref{Cetabeta1}) also by taking $1-\beta=1/T$ in \eqref{Ceta}, which is indeed the appropriate confidence level for maximal loss, because in a finite time window of $T$ points, the worst possible outcome occurs with probability $1-\beta=1/T$. 

\bibliography{IPO}{}
\bibliographystyle{ieeetr}
\end{document}